\RequirePackage[mathlines]{lineno} 
\documentclass[a4paper,11pt]{article}

\usepackage{jheppub} 

\usepackage[T1]{fontenc} 

\usepackage{threeparttable}
\usepackage{multirow}
\usepackage{subfigure}
\usepackage{float}

\usepackage{booktabs}

\let\oldequation\equation
\let\oldendequation\endequation
\renewenvironment{equation}
  {\linenomathNonumbers\oldequation}
  {\oldendequation\endlinenomath}

\def \gev  {~\mbox{GeV}}

\def \mev  {~\mbox{MeV}}

\def \mevcc{~\mbox{MeV/$c^2$}}

\begin{document}

\title{Could an axion-like particle be hidden in $\eta_c\to\gamma\gamma$?}

\author[a]{Zhi-Jun Li}
\author[a]{Zheng-Yun You}
\author[b]{Xiao-Rui Lyu}

\affiliation[a]{
School of Physics, Sun Yat-Sen University, Guangzhou 510275, China}
\affiliation[b]{
University of Chinese Academy of Sciences, Beijing 100049, China}

\emailAdd{lizhj37@mail2.sysu.edu.cn}
\emailAdd{youzhy5@mail.sysu.edu.cn}
\emailAdd{xiaorui@ucas.ac.cn}

\abstract{
We discuss the possibility of an axion-like particle being merged in the recent observation of $\eta_c \to \gamma\gamma$ at BESIII and find that it cannot be excluded in the current experimental data. The mass, width, and coupling strengths of the axion-like particle are extracted, and several other channels to confirm or exclude the existence of such an axion-like particle are also discussed.
}


\arxivnumber{xxxx.xxxx}

\maketitle
\flushbottom

\section{Introduction}
The axion is a new pseudoscalar gauge boson beyond the Standard Model (SM), predicted by the Peccei-Quinn solution to the strong CP problem~\cite{Peccei:1977hh,Peccei:1977ur,Weinberg:1977ma,Wilczek:1977pj}, and it was later applied to address the hierarchy problem~\cite{Graham:2015cka}. The axion-like particle (ALP) is an extension of the original axion, characterized by arbitrary masses and coupling strengths, interacting with SM photons and SM fermions through the following operator~\cite{Brivio:2017ije}:
\begin{eqnarray}
\mathcal{L}_{\rm{eff}}=-\frac{g_{a\gamma\gamma}}{4}aF_{\mu\nu}\tilde{F}^{\mu\nu}-\frac{g_{aff}}{2}\partial_{\mu}a\bar{f}\gamma^{\mu}\gamma_5f,
\end{eqnarray}
where $g_{a\gamma\gamma}$ ($g_{aff}$) is the coupling strength between the ALP and the SM photon (fermion). ALPs could serve as a portal between the dark sector and SM matter~\cite{Freytsis:2010ne} or act as a candidate for cold dark matter~\cite{Preskill:1982cy,Abbott:1982af,Dine:1982ah}. This has been predicted in several new physics (NP) models, such as supersymmetry~\cite{Bagger:1994hh}, extended Higgs sectors~\cite{Branco:2011iw}, and string theory~\cite{Witten:1984dg,Ringwald:2012cu}. However, no signals of ALPs have been found to date~\cite{Graham:2015ouw,Cadamuro:2011fd,NA64:2020qwq,BESIII:2020sdo,BaBar:2010eww,Belle:2018pzt,BaBar:2008aby,Knapen:2016moh,OPAL:2002vhf,L3:1994shn,Belle-II:2020jti,CMS:2018erd,ATLAS:2020hii,BESIII:2022rzz,BESIII:2024hdv}.

\begin{figure}[htbp] \centering
	\setlength{\abovecaptionskip}{-1pt}
	\setlength{\belowcaptionskip}{10pt}

        \subfigure[]
        {\includegraphics[width=0.41\textwidth]{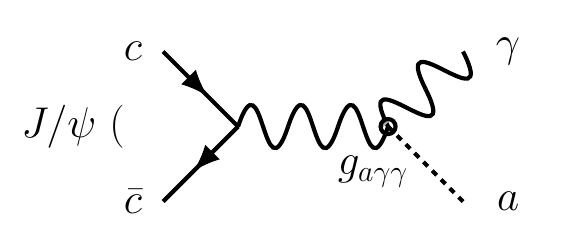}}
        \subfigure[]
        {\includegraphics[width=0.41\textwidth]{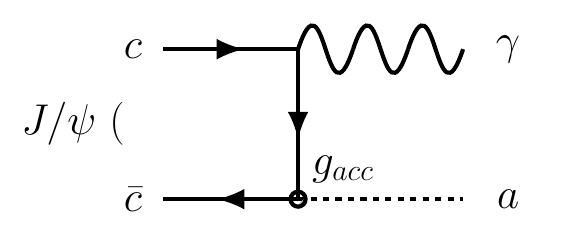}}\\
        
	\caption{The Feynman diagrams for $J/\psi\to\gamma a$ with the $g_{a\gamma\gamma}$ coupling (a) and $g_{acc}$ coupling (b).
    } 
	\label{fig:diagram}
\end{figure}

Recently, BESIII reported a direct measurement of $\eta_c \to \gamma\gamma$ in the decay channel $J/\psi \to \gamma \eta_c$, yielding the product branching fraction (BF) of $\mathcal{B}(J/\psi \to \gamma \eta_c) \times \mathcal{B}(\eta_c \to \gamma \gamma) = (5.23 \pm 0.40) \times 10^{-6}$~\cite{BESIII:2024rex}. 
By using the word-average value of $\mathcal{B}(J/\psi\to\gamma\eta_c)=(1.41\pm0.14)\%$, $\mathcal{B}(\eta_c \to \gamma \gamma)$ is calculated to be $(3.71\pm0.47)\times10^{-4}$, which is approximately one times larger than the value $(1.66\pm0.13)\times10^{-4}$ in the Particle Data Group (PDG).
Such a deviation is referred to as the $\eta_c \to \gamma\gamma$ anomaly.
The value of $\mathcal{B}(\eta_c \to \gamma \gamma)$ in PDG is predominantly determined by the time inversion process $\gamma \gamma \to \eta_c$ with $\eta_c \to \text{hadrons}$, making it an indirect measurement. To explain the discrepancies between the direct and indirect measurements, we propose the existence of an ALP that has a strong coupling with the charm quark or photon but a weak (or negligible) coupling with the light quarks. Such an ALP would have a small (or negligible) decay BF into hadrons, resulting in a minor contribution to the $\gamma\gamma \to \eta_c$ and $\eta_c \to \text{hadrons}$ measurements, but potentially a significant contribution to the $J/\psi \to \gamma \eta_c$ and $\eta_c \to \gamma \gamma$ measurements. The Feynman diagrams for the ALP produced in the decay process $J/\psi \to \gamma a$ are shown in Figure~\ref{fig:diagram}. 
In this paper, we will explore the possibility of the ALP decaying to $\gamma\gamma$ being hidden in the $\eta_c \to \gamma\gamma$ data from BESIII.

\vspace{-0.0cm}
\vspace{-0.0cm}

\section{Re-fit the BESIII data}
\begin{figure}[htbp] \centering
	\setlength{\abovecaptionskip}{-1pt}
	\setlength{\belowcaptionskip}{10pt}

        {\includegraphics[width=0.8\textwidth]{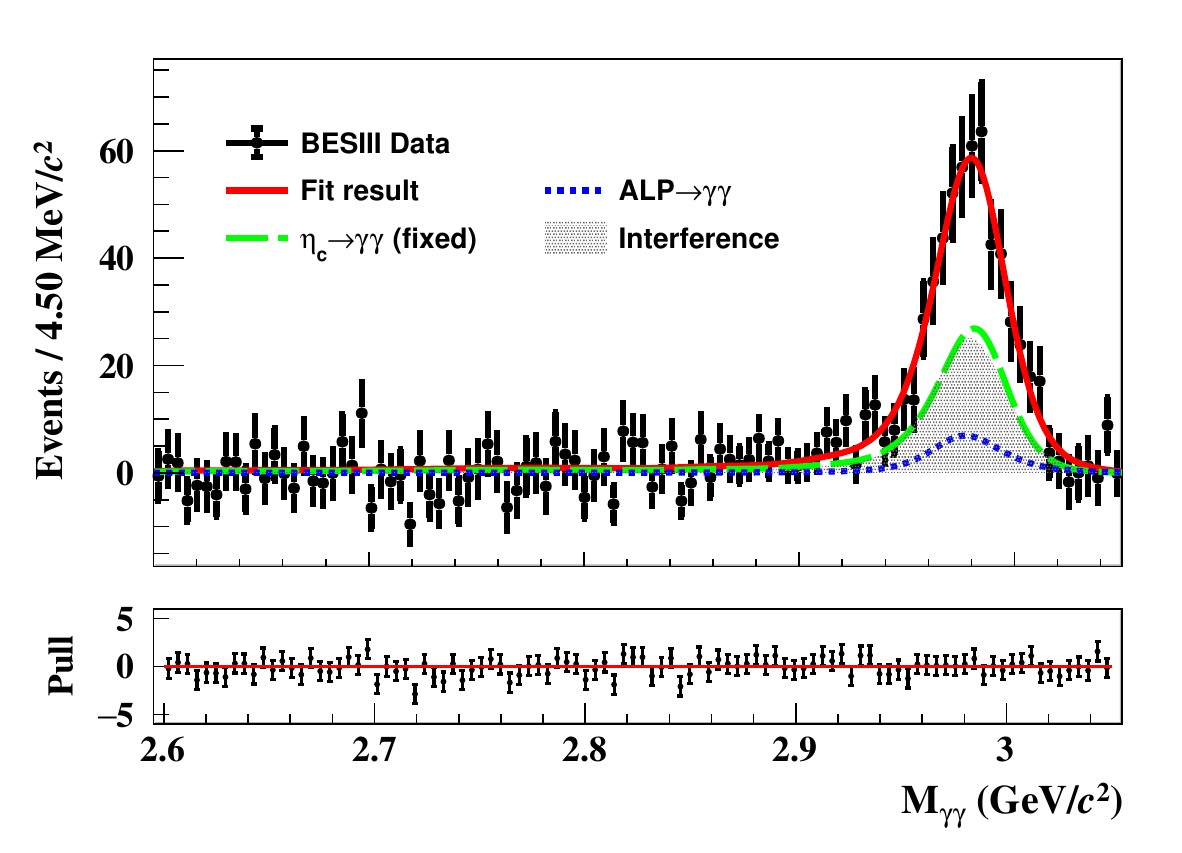}}
        
	\caption{
        Fit to the $M_{\gamma\gamma}$ distribution with $\phi=0$. The black points with error bars represent the BESIII data samples after removing the background component. The red line indicates the fit result, the green dashed line corresponds to $\eta_c \to \gamma\gamma$, the blue dashed line represents $a \to \gamma\gamma$, and the gray-filled histogram illustrates the interference between $\eta_c$ and the ALP.
    } 
	\label{fig:fit}
\end{figure}

To extract the signal yield, mass, and width of our ALP, we refit the invariant mass distribution of di-photons ($M_{\gamma\gamma}$) from BESIII~\cite{BESIII:2024rex} by introducing an ALP resonance. The background component is directly removed following the BESIII study~\cite{BESIII:2024rex} during the fitting process. The fitting probability density function is given by:
\begin{equation}
\begin{split}
    \mathcal{PDF}(m)\sim[\epsilon(m)\times|e^{i\phi}\sqrt{E^3_{\gamma}(m)f_{\rm{damp}}(m)}\times BW_{\eta_c}(m)
+\alpha BW_{a}(m)|^2]\otimes G(\mu,\sigma),
\end{split}
\end{equation}
where $m$ denotes $M_{\gamma\gamma}$, $\epsilon(m)$ is the mass-dependent efficiency, $E^3_{\gamma}(m)$ represents the M1 transition form factor, $f_{\rm{damp}}(m)$ is the damping factor, $G(\mu, \sigma)$ is the convoluted Gaussian function, and $BW_{\eta_c}(m)$ is the Breit-Wigner function for $\eta_c$, all of which are fixed according to the BESIII paper~\cite{BESIII:2024rex}. The parameter $\alpha$ denotes the strength of the ALP contribution, $\phi$ is the interference phase, and $BW_a(m)$ is the Breit-Wigner function for the ALP, with
\begin{eqnarray}
BW_a(m)=\frac{m_{a}\Gamma_{a}}{m^2-m_{a}^2+im_{a}\Gamma_{a}},
\end{eqnarray}
where $m_a$ is the mass of the ALP, and $\Gamma_a$ is the width of the ALP. In the fit, we assume that the measurements from $\gamma\gamma \to \eta_c$ and $\eta_c \to \text{hadrons}$ are based on a pure $\eta_c$ state, fixing the contribution of $\eta_c \to \gamma\gamma$ using the world-average value $\mathcal{B}(J/\psi \to \gamma \eta_c) \times \mathcal{B}(\eta_c \to \gamma \gamma)$ in the current PDG. The uncertainty of the PDG value will be considered later. Different values of $\phi$ in the fit will produce either constructive or destructive interference between $\eta_c$ and the ALP. We find that the current BESIII dataset is unable to distinguish between different $\phi$ values, as they produce similar fitting quality. To estimate a minimum ALP contribution needed to explain the $\eta_c \to \gamma\gamma$ anomaly, we fix $\phi=0$ in the fit, which maximizes constructive interference. The fitting result is shown in Figure~\ref{fig:fit} with $N_{a \to \gamma \gamma} = 73.3^{+9.7}_{-8.4}$. The mass and width of the ALP are determined to be
\begin{eqnarray}
m_a=(2977.5\pm2.6)\mevcc
\label{eq:exp_alp_mass}
\end{eqnarray}
and
\begin{eqnarray}
\Gamma_a=(31.3\pm6.7)\mev.
\label{eq:exp_alp_width}
\end{eqnarray}
The uncertainty in the $\eta_c$ contribution is accounted for by varying the fixed yield within the standard deviation of $\mathcal{B}(J/\psi \to \gamma \eta_c) \times \mathcal{B}(\eta_c \to \gamma \gamma)$ from the PDG, and the resulting changes are considered in the uncertainty of the ALP. Ultimately, $m_a$ and $\Gamma_a$ remain stable, and $N_{a \to \gamma \gamma} = 73.3^{+23.5}_{-19.7}$. Taking the $2\sigma$ lower bound of $N_{a \to \gamma \gamma}$, to explain the $\eta_c \to \gamma\gamma$ anomaly, there would be
\begin{eqnarray}
\mathcal{B}(J/\psi\to\gamma a)\times\mathcal{B}(a\to\gamma\gamma)>2.6\times10^{-7}.
\label{eq:exp_BF_Jpsi2ga2ggg}
\end{eqnarray}
Here, the BF is calculated similarly to the method described in the BESIII paper~\cite{BESIII:2024rex}.

\vspace{-0.0cm}
\vspace{-0.0cm}

\section{Coupling with the SM particles}
Next, we will discuss the current constraints on the ALP and explore whether such an ALP can exist.

Considering the non-zero couplings of $g_{a\gamma\gamma}$ and $g_{acc}$ in Figure~\ref{fig:diagram}, the BF of $J/\psi\to\gamma a$ can be expressed as~\cite{Merlo:2019anv,DiLuzio:2024jip}:
\begin{equation}
\begin{split}
    \mathcal{B}(J/\psi\to\gamma a) = \frac{\alpha_{\rm{em}}}{216\Gamma_{J/\psi}} m_{J/\psi} f^2_{J/\psi} \left(1-\frac{m^2_a}{m^2_{J/\psi}}\right) 
     \times \left[ g_{a\gamma\gamma}(1-\frac{m^2_a}{m^2_{J/\psi}}) - 2g_{acc} \right]^2,
    \label{eq:the_BF_Jpsi2ga}
\end{split}
\end{equation}
where $\alpha_{\rm{em}}$ is the fine structure constant, $m_{J/\psi}$ ($\Gamma_{J/\psi}$) is the mass (total width) of $J/\psi$~\cite{pdg:2024}, and $f_{J/\psi}$ is the decay constant of $J/\psi$, calculated using the experimental value of $\Gamma_{J/\psi \to e^+ e^-}$~\cite{pdg:2024,Merlo:2019anv,Becirevic:2013bsa}. 
Note that Eq~(\ref{eq:the_BF_Jpsi2ga}) is derived solely from the tree-level contribution, and further $\mathcal{O}(\alpha_s)$ NRQCD perturbative analyses~\cite{Nason:1986tr,Bauer:2021mvw,Carmona:2021seb} as well as the first fully non-perturbative analysis from lattice QCD~\cite{Colquhoun:2025xlx} have been conducted. 
These further corrections have little impact on $\mathcal{B}(J/\psi \to \gamma a)$ when $m_a$ is around 2.98 GeV~\cite{Colquhoun:2025xlx}, and we only use the tree-level contribution in this paper to simplify our discussion. A sizable coupling strength of $g_{a\gamma\gamma}$ and $g_{acc}$ can account for the observed BF in Eq~(\ref{eq:exp_BF_Jpsi2ga2ggg}) from BESIII.

For the current constraint on $g_{acc}$, it can only be obtained from charm quark decays.
There are several searches for $J/\psi \to \gamma a$ with $a \to \text{invisible}$~\cite{BESIII:2020sdo} (we will return to the discussion of ALP invisible decay later) and for $J/\psi \to \gamma a$ with $a \to \gamma\gamma$~\cite{BESIII:2022rzz,BESIII:2024hdv}. However, these searches have not reached the mass of the ALP in Eq~(\ref{eq:exp_alp_mass}), and thus cannot provide an effective constraint for our scenario. Another loose constraint may arise from the measured channels of $J/\psi$. By summing the BF of hadronic and leptonic decays of $J/\psi$, we have $1 - \mathcal{B}_{\rm{H+L}}(J/\psi) = (0.4 \pm 0.5)\%$~\cite{pdg:2024}. Taking the $2\sigma$ upper bound, we simply have
\begin{eqnarray}
\mathcal{B}(J/\psi\to\gamma a)<1.4\%.
\label{eq:exp_BF_Jpsi2ga}
\end{eqnarray}

For the current constraint on $g_{a\gamma\gamma}$, an important pure constraint should come from the process $e^+ e^- \to \gamma a$ with $a \to \gamma \gamma$, where both couplings are related to the pure $g_{a\gamma\gamma}$ in the production and decay of the ALP (it should be noted that the coupling between the ALP and the electron is not considered in this analysis). In previous experimental studies of $e^+ e^- \to \gamma a$ with $a \to \gamma \gamma$, the nature of the ALP is ambiguous, and the BF of $a \to \gamma \gamma$ was always assumed to be 100\% to provide the constraint on $g'_{a\gamma\gamma}$. 
If we consider a non-100\% BF for $a \to \gamma \gamma$, the true constraint on $g_{a\gamma\gamma}$ should be modified to $g_{a\gamma\gamma} = \sqrt{\frac{g'^2_{a\gamma\gamma}}{\mathcal{B}(a \to \gamma \gamma)}}$. 
On the theoretical side, the BF of $a \to \gamma \gamma$ is expressed as~\cite{Dolan:2017osp}:
\begin{eqnarray}
\mathcal{B}(a\to\gamma\gamma)=\frac{g^2_{a\gamma\gamma}m^3_a}{64\pi\Gamma_a},
\label{eq:the_BF_a2gg}
\end{eqnarray}
where the ALP total width $\Gamma_a$ has been obtained in Eq~(\ref{eq:exp_alp_width}). The most direct and stringent constraint on the process $e^+ e^- \to \gamma a$ with $a \to \gamma \gamma$ comes from Belle II, which provides the limit $g'_{a\gamma\gamma} < 2.3 \times 10^{-3} \ \text{GeV}^{-1}$ for $m_a$ as given in Eq~(\ref{eq:exp_alp_mass})~\cite{Belle-II:2020jti}. Considering the BF of $a \to \gamma \gamma$ and using Eqs~(\ref{eq:the_BF_a2gg}) and (\ref{eq:exp_alp_width}), we can obtain:
\begin{eqnarray}
\mathcal{B}(a\to\gamma\gamma)<4.7\times10^{-3}.
\label{eq:exp_BF_a2gg_up}
\end{eqnarray}
We should note that there could be another slightly ``better" constraint from OPAL's $e^+ e^- \to \gamma \gamma$ data at $\sqrt{s} \sim 200 \ \text{GeV}$~\cite{OPAL:2002vhf}. This constraint has been extracted by Ref.~\cite{Knapen:2016moh}, yielding $g'_{a\gamma\gamma} < 1.3 \times 10^{-3} \ \text{GeV}^{-1}$ for ALP masses less than 8 GeV. 
For such small ALP masses compared to $\sqrt{s}$, the photon pair from the ALP decay is collimated and may be misidentified as a single photon, thus appearing as part of the $e^+ e^- \to \gamma \gamma$ final state. Ref.~\cite{Knapen:2016moh} simply estimated the selection efficiency for this process using MadGraph and subsequently calculated the constraint using OPAL's $e^+ e^- \to \gamma \gamma$ data, which may lead to an overestimation of both the efficiency and the constraint. Considering that the constraint from Ref.~\cite{Knapen:2016moh} is not significantly better than that from Belle II~\cite{Belle-II:2020jti}, we prefer to use the upper limit from Belle II for our discussion.

\begin{figure}[htbp] \centering
	\setlength{\abovecaptionskip}{-1pt}
	\setlength{\belowcaptionskip}{10pt}

        {\includegraphics[width=0.8\textwidth]{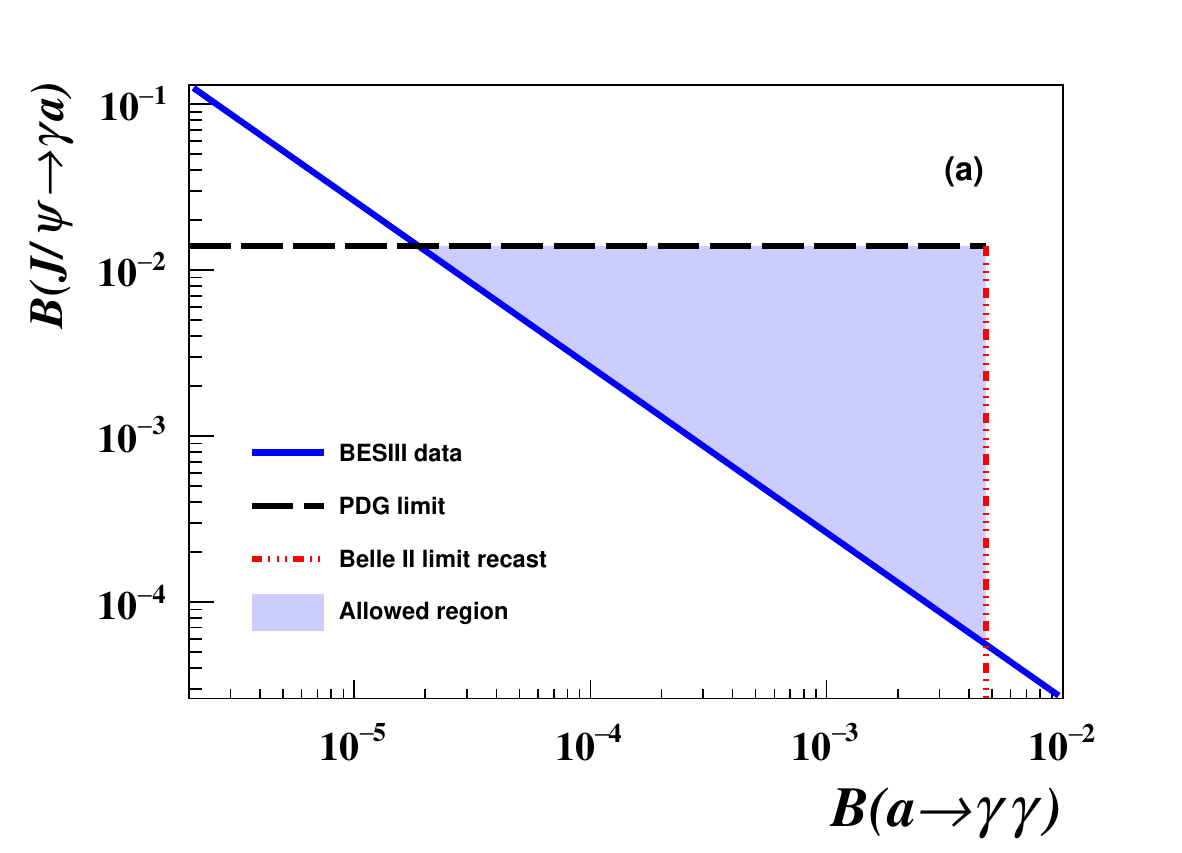}}\\
        {\includegraphics[width=0.8\textwidth]{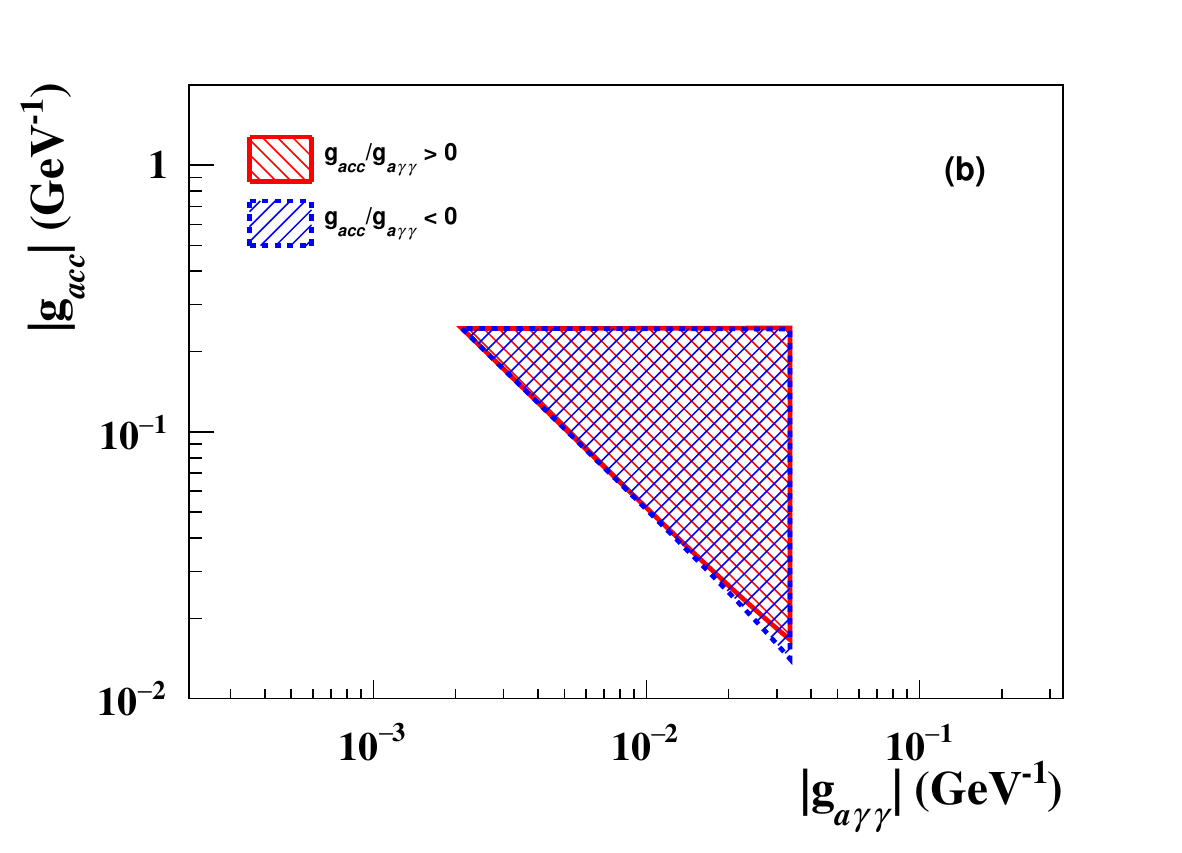}}\\
        
	\caption{
        (a) The $J/\psi \to \gamma a$ versus $a \to \gamma\gamma$ BF space. The blue solid line represents the ALP BF derived from BESIII data, the black dashed line indicates the limit from PDG~\cite{pdg:2024}, the red dashed line shows the recast limit from Belle II~\cite{Belle-II:2020jti}, and the blue shaded region denotes the allowed region.
        (b) The corresponding parameter space for $|g_{acc}|$ and $|g_{a\gamma\gamma}|$ to explain the $\eta_c \to \gamma\gamma$ anomaly. The red shaded region corresponds to the case where $\frac{g_{acc}}{g_{a\gamma\gamma}} > 0$, while the blue shaded region corresponds to the case where $\frac{g_{acc}}{g_{a\gamma\gamma}} < 0$.
    }
	\label{fig:BF_coupling}
\end{figure}

Combining Eqs~(\ref{eq:exp_BF_Jpsi2ga2ggg}), (\ref{eq:exp_BF_Jpsi2ga}), and (\ref{eq:exp_BF_a2gg_up}), the BF space of $J/\psi \to \gamma a$ and $a \to \gamma \gamma$ is shown in Figure~\ref{fig:BF_coupling} (a). Using Eqs~(\ref{eq:the_BF_Jpsi2ga}) and (\ref{eq:the_BF_a2gg}), the corresponding parameter space of $|g_{acc}|$ and $|g_{a\gamma\gamma}|$ is illustrated in Figure~\ref{fig:BF_coupling} (b), considering both scenarios of $\frac{g_{acc}}{g_{a\gamma\gamma}} > 0$ and $\frac{g_{acc}}{g_{a\gamma\gamma}} < 0$ in Eq~(\ref{eq:the_BF_Jpsi2ga}).
The two scenarios show a small difference due to the contribution of $g_{a\gamma\gamma}$ to $\mathcal{B}(J/\psi \to \gamma a)$ being suppressed by a factor of $1 - \frac{m_a^2}{m_{J/\psi}^2}$ when $m_a$ is close to $m_{J/\psi}$. 
From Figure~\ref{fig:BF_coupling}, it is evident that the current experimental searches are unable to rule out the existence of our ALP as a potential explanation for the $\eta_c \to \gamma\gamma$ anomaly.

\vspace{-0.0cm}
\vspace{-0.0cm}

\section{Coupling with the dark sector}
\begin{figure}[htbp] \centering
	\setlength{\abovecaptionskip}{-1pt}
	\setlength{\belowcaptionskip}{10pt}

        {\includegraphics[width=0.8\textwidth]{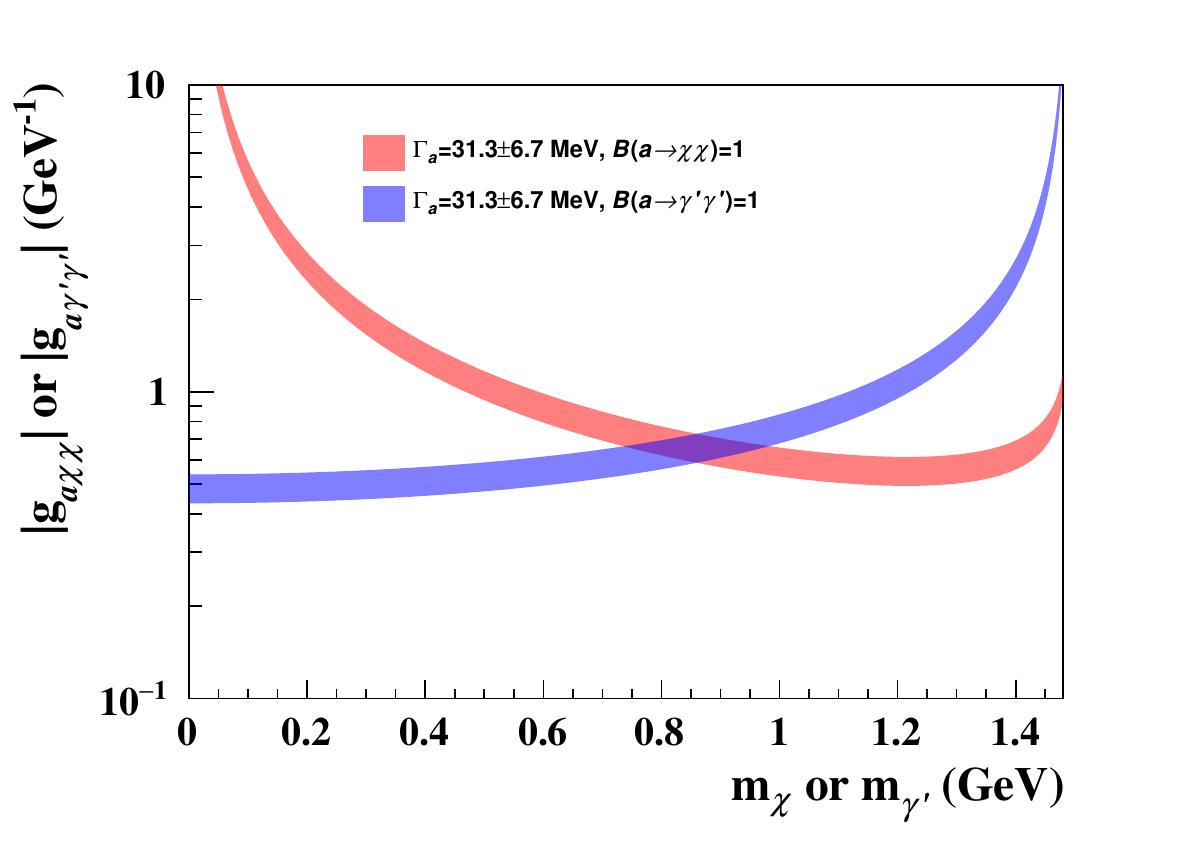}}
        
	\caption{
        The allowed parameter space for $g_{a\chi\chi}-m_{\chi}$ ($g_{a\gamma'\gamma'}-m_{\gamma'}$) to account for the observed width of the ALP. The red-filled region corresponds to the assumption of $\mathcal{B}(a\to\chi\chi)=1$, while the blue-filled region represents the assumption of $\mathcal{B}(a\to\gamma'\gamma')=1$.
    } 
	\label{fig:DM}
\end{figure}

However, the small BF of $a\to\gamma\gamma$ in Figure~\ref{fig:BF_coupling} (a) cannot account for the large total width of the ALP obtained from BESIII in Eq~(\ref{eq:exp_alp_width}). To address this discrepancy, the ALP should decay into other final states. As discussed in the introduction of ALP in the first paragraph, the ALP can serve as a portal between the dark sector and the SM matter~\cite{Freytsis:2010ne}. In this scenario, new couplings are typically introduced to describe these additional interactions. We assume the existence of a dark fermion $\chi$ or a dark photon $\gamma'$, and the ALP can interact with these dark particles through the similar operator:
\begin{eqnarray}
\mathcal{L}'_{\rm{eff}}=-\frac{g_{a\chi\chi}}{2}\partial_{\mu}a\bar{\chi}\gamma^{\mu}\gamma_5\chi-\frac{g_{a\gamma'\gamma'}}{4}aF'_{\mu\nu}\tilde{F'}^{\mu\nu}.
\end{eqnarray}
If the mass of $\chi$ or $\gamma'$ is less than half of the mass of the ALP, the decay width of the ALP would be~\cite{Allen:2024ndv}:
\begin{eqnarray}
\Gamma_{a\to\chi\bar{\chi}}=\frac{g^2_{a\chi\chi}m_am^2_{\chi}}{8\pi}\sqrt{1-\frac{4m^2_{\chi}}{m^2_a}}
\end{eqnarray}
and~\cite{Kaneta:2016wvf} 
\begin{eqnarray}
\Gamma_{a\to\gamma'\gamma'}=\frac{g^2_{a\gamma'\gamma'}m^3_a}{64\pi}\left[1-\frac{4m^2_{\gamma'}}{m^2_a}\right]^{3/2}
\label{eq:W_a2gpgp}
\end{eqnarray}
for $a\to\chi\chi$ and $a\to\gamma'\gamma'$ respectively,  
where $g_{a\chi\chi}$ ($g_{a\gamma'\gamma'}$) is the coupling strength between the ALP and the dark fermion (dark photon).  
To satisfy the width observed in Eq~(\ref{eq:exp_alp_width}), the coupling strength $|g_{a\chi\chi}|$ ($|g_{a\gamma'\gamma'}|$) should lie within a specific range, as illustrated in Figure~\ref{fig:DM}. In the plot, the BF of $a\to\chi\chi$ ($a\to\gamma'\gamma'$) is assumed to be 1, and one can easily derive the corresponding coupling strength using $g^2\sim\rm{BF}$.

\vspace{-0.0cm}
\vspace{-0.0cm}

If the unknown decay final state of the ALP is dark matter (DM), it is interesting to explore whether the parameter space of our ALP can reproduce the observed DM relic abundance. Here, we refrain from delving into this topic in depth, as it heavily depends on the specific DM model, and we leave it for future study. For simplicity, we consider $\chi$ as the DM and follow the formula provided in Ref.~\cite{Dolan:2017osp}. Assuming $\chi\chi$ annihilates into $\gamma\gamma$ through the ALP mediator, if the cross section for $\chi\chi\to\gamma\gamma$ is close to the thermal value, $(\sigma v)_{\rm{th}}\simeq3\times10^{-26}~\rm{cm}^3/s$, the DM particle can, in principle, attain its relic abundance via thermal freeze-out. The annihilation cross section is given by~\cite{Dolan:2017osp}:
\begin{eqnarray}
\sigma(\chi\chi\to\gamma\gamma)v\simeq\frac{g^2_{a\gamma\gamma}g^2_{a\chi\chi}m^6_{\chi}}{\pi m^4_a}.
\end{eqnarray}
Taking $g_{a\gamma\gamma}=10^{-2}\gev^{-1}$, $m_{\chi}=0.4\gev$, and $g_{a\chi\chi}=1.5\gev^{-1}$ as the example, which are allowed by the parameter spaces shown in Figure~\ref{fig:BF_coupling} (b) and Figure~\ref{fig:DM}, the resulting cross section is $\sigma v\simeq4\times10^{-26}~\rm{cm}^3/s$, which is comparable to the thermal cross section $(\sigma v)_{\rm{th}}$. This is a preliminary estimation, and a more detailed investigation is warranted.

Since the final dark particles from ALP decay could be invisible in the final states, we now reconsider the constraint on $g_{a\gamma\gamma}$ from the measurement of ALP invisible decays.  
Unfortunately, there is no direct search for $J/\psi\to\gamma a$ or $e^+e^-\to\gamma a$ with $a\to\rm{invisible}$ for the ALP mass in Eq~(\ref{eq:exp_alp_mass}).  
One related search is $e^+e^-\to\gamma\gamma'$, $\gamma'\to\rm{invisible}$ from BaBar~\cite{BaBar:2017tiz}, but the kinematic distributions differ for spin-0 ALP and spin-1 dark photon scenarios, and the detector efficiency also varies between the two cases. Ref.~\cite{Dolan:2017osp} performed a reinterpretation of the dark photon data from BaBar without accounting for the efficiency difference, which may overestimate the limit~\cite{Merlo:2019anv}.  
Another related search is $\Upsilon(nS)\to\gamma a$, $a\to\rm{invisible}$~\cite{BaBar:2010eww,Belle:2018pzt,BaBar:2008aby}. For such a process, its BF of ALP production should be similar to Eq~(\ref{eq:the_BF_Jpsi2ga}), and the measurement is sensitive to $|g_{a\gamma\gamma}(1-\frac{m^2_a}{m^2_\Upsilon})-2g_{abb}|$. If $g_{a\gamma\gamma}$ shares the same sign as $g_{abb}$, there would be a linear gap in the $g_{a\gamma\gamma}-g_{abb}$ space that cannot be reached, making the constraint inapplicable directly to $J/\psi\to\gamma a$.
In addition, the result of $\Upsilon(3S)\to\gamma a$, $a\to\rm{invisible}$ in 2008 ~\cite{BaBar:2008aby} remains a conference result and has not been officially published to date.
To avoid potential biases that might lead us to overlook the ALP, we prefer to await a dedicated search for $e^+e^-\to\gamma a$ with $a\to\rm{invisible}$ without resonance contributions from the BaBar, BESIII, or Belle II collaborations in the future.

\section{Phenomenon}
Next, we will explore additional channels that could either confirm or exclude the existence of our proposed ALP.

The first important verification channel should be $\psi(2S)\to\gamma a$, $a\to\gamma\gamma$ at BESIII. The BF of $\psi(2S)\to\gamma a$ is similar to Eq~(\ref{eq:the_BF_Jpsi2ga}) but with the label of $J/\psi$ replaced by $\psi(2S)$. Using the BF of $J/\psi\to\gamma a$, $a\to\gamma\gamma$ in Eq~(\ref{eq:exp_BF_Jpsi2ga}), $\mathcal{B}(\psi(2S)\to\gamma a)\times\mathcal{B}(a\to\gamma\gamma)$ is predicted to be larger than $2.3\times10^{-7}$.  
Comparing this with $\mathcal{B}(\psi(2S)\to\gamma \eta_c)\times\mathcal{B}(\eta_c\to\gamma\gamma)=(6.1\pm1.0)\times10^{-7}$ from the PDG, the ALP is expected to contribute a significant observable effect.  
Additionally, the cross section of $e^+e^-\to\gamma a$, $a\to\gamma\gamma$ can be expressed as:
\begin{eqnarray}
\sigma(e^+e^-\to\gamma(\gamma\gamma)_a)=\frac{\alpha_{\rm{em}}g^4_{a\gamma\gamma}m^3_a}{1536\pi\Gamma_a}\left(1-\frac{m^2_a}{s}\right)^3,
\end{eqnarray}
where $\sqrt{s}$ is the center-of-mass energy of $e^+e^-$. Using the parameter space of $g_{a\gamma\gamma}$ in Figure~\ref{fig:BF_coupling} (b), the cross section is expected to be $\sigma(e^+e^-\to\gamma(\gamma\gamma)_a)\sim[1.0\times10^{-8},6.4\times10^{-4}]\times(1-2977.5~\mev^2/s)^3~\rm{nb}$, which could be validated in future experiments such as future super tau-charm facility (STCF)~\cite{Achasov:2023gey} and Belle II~\cite{Belle-II:2010dht}.

Other than $a \to \gamma \gamma$, $a \to \text{dark~sector}$ is expected to have a much larger BF ($\sim 100\%$), for example, ALP decays to invisible final states (but we cannot conclude that $\mathcal{B}(a \to \text{invisible})$ must be 100\%). 
Based on the BF space in Figure~\ref{fig:BF_coupling} (a), $\mathcal{B}(J/\psi \to \gamma a)$ is predicted to be in the region of $(5.6 \times 10^{-5}, \, 1.4 \times 10^{-2})$, and the corresponding $\mathcal{B}(\psi(2S) \to \gamma a)$ should be in the region of $(4.8 \times 10^{-6}, \, 1.2 \times 10^{-2})$. BESIII has collected $\sim 1 \times 10^{10}$ $J/\psi$ events~\cite{BESIII:2021cxx} and $2.7 \times 10^{9}$ $\psi(2S)$ events~\cite{BESIII:2017tvm,BESIII:2024lks}, and STCF is expected to collect $\sim$3 trillion $J/\psi$ events and $\sim$1 trillion $\psi(2S)$ events~\cite{Achasov:2023gey}. If $\mathcal{B}(a \to \text{invisible})$ is large, this should be within the sensitivity of BESIII or the future super tau-charm facility.
The future direct search for $e^+ e^- \to \gamma a$, $a \to \text{invisible}$ is also crucial. Ref.~\cite{Dolan:2017osp} has studied the expected sensitivity of this process at Belle II and found that the sensitivity can reach below $g_{a\gamma\gamma} < 10^{-3} \ \text{GeV}^{-1}$ based on the assumption of $\mathcal{B}(a \to \text{invisible}) = 1$, which can either confirm or exclude our ALP shown in Figure~\ref{fig:BF_coupling} (b). 
However, it should be noted again that we do not actually know the true value of $\mathcal{B}(a \to \text{invisible})$, and this uncertainty would impact the experimental sensitivity. For instance, considering the decay $a \to \gamma' \gamma'$ in Eq~(\ref{eq:W_a2gpgp}), where the dark photon has a kinetic mixing with the SM photon and subsequently decays to SM fermions such as $\gamma' \to e^+ e^-$ or $\gamma' \to \mu^+ \mu^-$~\cite{Holdom:1985ag,Buschmann:2015awa}, the dark sector decay channels of the ALP would not be invisible. Furthermore, the ALP can also decay to semi-dark sector final states, for example, $a \to \gamma \gamma'$ through the operator~\cite{Kaneta:2016wvf}.
\begin{eqnarray}
\mathcal{L}''_{\rm{eff}}=-\frac{g_{a\gamma\gamma'}}{2}aF_{\mu\nu}\tilde{F'}^{\mu\nu}
\end{eqnarray}
with the new coupling strength $g_{a\gamma\gamma'}$, leading to the decay width of~\cite{Kaneta:2016wvf}.
\begin{eqnarray}
\Gamma_{a\to\gamma\gamma'}=\frac{g^2_{a\gamma\gamma'}m^3_a}{32\pi}\left[1-\frac{m^2_{\gamma'}}{m^2_a}\right]^{3/2}.
\end{eqnarray}
In other words, in the search for ALPs, some other reconstructed channels such as $a \to l^+ l^- l'^+ l'^-$, $a \to \gamma \, \text{invisible}$, and $a \to \gamma l^+ l^-$ (where $l$ denotes the SM lepton) are also crucial. These channels have always been overlooked in previous searches.

\section{Summary}
In summary, to explain the observed difference between $J/\psi \to \gamma \eta_c$, $\eta_c \to \gamma \gamma$ and $\gamma \gamma \to \eta_c$, $\eta_c \to \text{hadrons}$, we propose that there may be an ALP generated from $J/\psi \to \gamma a$ that is hidden in the $\eta_c \to \gamma \gamma$ decays. Such an ALP should have the following characteristics:
\begin{itemize}
    \item It has small couplings with the light quarks.
    \item Its mass and width are similar to those of $\eta_c$ (see Eq~(\ref{eq:exp_alp_mass}) and Eq~(\ref{eq:exp_alp_width})).
    \item It has sizable couplings with the charm quark and photons (see Figure~\ref{fig:BF_coupling} (b)).
    \item It has large couplings with other light dark sector particles (see Figure~\ref{fig:DM}).
\end{itemize}
We find that such an ALP may not be excluded by the current experimental searches. Future experimental searches for $J/\psi \to \gamma a$, $\psi(2S) \to \gamma a$, $e^+ e^- \to \gamma a$ with $a \to \gamma \gamma$, $a \to \text{invisible}$, $a \to l^+ l^- l'^+ l'^-$, $a \to \gamma \, \text{invisible}$, and $a \to \gamma l^+l^-$ are expected to confirm or exclude the possibility of this ALP.

\acknowledgments
\hspace{1.5em}
This work is supported by National Key R\&D Program of China under Contracts Nos. 2023YFA1606000.

\bibliographystyle{JHEP.bst} 
\bibliography{mybib.bib}

\providecommand{\href}[2]{#2}\begingroup\raggedright\begin{thebibliography}{10}

\bibitem{Peccei:1977hh}
R.D.~Peccei and H.R.~Quinn, \emph{{CP Conservation in the Presence of
  Instantons}}, \href{https://doi.org/10.1103/PhysRevLett.38.1440}{\emph{Phys.
  Rev. Lett.} {\bfseries 38} (1977) 1440}.

\bibitem{Peccei:1977ur}
R.D.~Peccei and H.R.~Quinn, \emph{{Constraints Imposed by CP Conservation in
  the Presence of Instantons}},
  \href{https://doi.org/10.1103/PhysRevD.16.1791}{\emph{Phys. Rev. D}
  {\bfseries 16} (1977) 1791}.

\bibitem{Weinberg:1977ma}
S.~Weinberg, \emph{{A New Light Boson?}},
  \href{https://doi.org/10.1103/PhysRevLett.40.223}{\emph{Phys. Rev. Lett.}
  {\bfseries 40} (1978) 223}.

\bibitem{Wilczek:1977pj}
F.~Wilczek, \emph{{Problem of Strong $P$ and $T$ Invariance in the Presence of
  Instantons}}, \href{https://doi.org/10.1103/PhysRevLett.40.279}{\emph{Phys.
  Rev. Lett.} {\bfseries 40} (1978) 279}.

\bibitem{Graham:2015cka}
P.W.~Graham, D.E.~Kaplan and S.~Rajendran, \emph{{Cosmological Relaxation of
  the Electroweak Scale}},
  \href{https://doi.org/10.1103/PhysRevLett.115.221801}{\emph{Phys. Rev. Lett.}
  {\bfseries 115} (2015) 221801}
  [\href{https://arxiv.org/abs/1504.07551}{{\ttfamily 1504.07551}}].

\bibitem{Brivio:2017ije}
I.~Brivio, M.B.~Gavela, L.~Merlo, K.~Mimasu, J.M.~No, R.~del Rey et~al.,
  \emph{{ALPs Effective Field Theory and Collider Signatures}},
  \href{https://doi.org/10.1140/epjc/s10052-017-5111-3}{\emph{Eur. Phys. J. C}
  {\bfseries 77} (2017) 572}
  [\href{https://arxiv.org/abs/1701.05379}{{\ttfamily 1701.05379}}].

\bibitem{Freytsis:2010ne}
M.~Freytsis and Z.~Ligeti, \emph{{On dark matter models with uniquely
  spin-dependent detection possibilities}},
  \href{https://doi.org/10.1103/PhysRevD.83.115009}{\emph{Phys. Rev. D}
  {\bfseries 83} (2011) 115009}
  [\href{https://arxiv.org/abs/1012.5317}{{\ttfamily 1012.5317}}].

\bibitem{Preskill:1982cy}
J.~Preskill, M.B.~Wise and F.~Wilczek, \emph{{Cosmology of the Invisible
  Axion}}, \href{https://doi.org/10.1016/0370-2693(83)90637-8}{\emph{Phys.
  Lett. B} {\bfseries 120} (1983) 127}.

\bibitem{Abbott:1982af}
L.F.~Abbott and P.~Sikivie, \emph{{A Cosmological Bound on the Invisible
  Axion}}, \href{https://doi.org/10.1016/0370-2693(83)90638-X}{\emph{Phys.
  Lett. B} {\bfseries 120} (1983) 133}.

\bibitem{Dine:1982ah}
M.~Dine and W.~Fischler, \emph{{The Not So Harmless Axion}},
  \href{https://doi.org/10.1016/0370-2693(83)90639-1}{\emph{Phys. Lett. B}
  {\bfseries 120} (1983) 137}.

\bibitem{Bagger:1994hh}
J.~Bagger, E.~Poppitz and L.~Randall, \emph{{The R axion from dynamical
  supersymmetry breaking}},
  \href{https://doi.org/10.1016/0550-3213(94)90123-6}{\emph{Nucl. Phys. B}
  {\bfseries 426} (1994) 3}
  [\href{https://arxiv.org/abs/hep-ph/9405345}{{\ttfamily hep-ph/9405345}}].

\bibitem{Branco:2011iw}
G.C.~Branco, P.M.~Ferreira, L.~Lavoura, M.N.~Rebelo, M.~Sher and J.P.~Silva,
  \emph{{Theory and phenomenology of two-Higgs-doublet models}},
  \href{https://doi.org/10.1016/j.physrep.2012.02.002}{\emph{Phys. Rept.}
  {\bfseries 516} (2012) 1} [\href{https://arxiv.org/abs/1106.0034}{{\ttfamily
  1106.0034}}].

\bibitem{Witten:1984dg}
E.~Witten, \emph{{Some Properties of O(32) Superstrings}},
  \href{https://doi.org/10.1016/0370-2693(84)90422-2}{\emph{Phys. Lett. B}
  {\bfseries 149} (1984) 351}.

\bibitem{Ringwald:2012cu}
A.~Ringwald, \emph{{Searching for axions and ALPs from string theory}},
  \href{https://doi.org/10.1088/1742-6596/485/1/012013}{\emph{J. Phys. Conf.
  Ser.} {\bfseries 485} (2014) 012013}
  [\href{https://arxiv.org/abs/1209.2299}{{\ttfamily 1209.2299}}].

\bibitem{Graham:2015ouw}
P.W.~Graham, I.G.~Irastorza, S.K.~Lamoreaux, A.~Lindner and K.A.~van Bibber,
  \emph{{Experimental Searches for the Axion and Axion-Like Particles}},
  \href{https://doi.org/10.1146/annurev-nucl-102014-022120}{\emph{Ann. Rev.
  Nucl. Part. Sci.} {\bfseries 65} (2015) 485}
  [\href{https://arxiv.org/abs/1602.00039}{{\ttfamily 1602.00039}}].

\bibitem{Cadamuro:2011fd}
D.~Cadamuro and J.~Redondo, \emph{{Cosmological bounds on pseudo
  Nambu-Goldstone bosons}},
  \href{https://doi.org/10.1088/1475-7516/2012/02/032}{\emph{JCAP} {\bfseries
  02} (2012) 032} [\href{https://arxiv.org/abs/1110.2895}{{\ttfamily
  1110.2895}}].

\bibitem{NA64:2020qwq}
{\scshape NA64} collaboration, \emph{{Search for Axionlike and Scalar Particles
  with the NA64 Experiment}},
  \href{https://doi.org/10.1103/PhysRevLett.125.081801}{\emph{Phys. Rev. Lett.}
  {\bfseries 125} (2020) 081801}
  [\href{https://arxiv.org/abs/2005.02710}{{\ttfamily 2005.02710}}].

\bibitem{BESIII:2020sdo}
{\scshape BESIII} collaboration, \emph{{Search for the decay $J/\psi\to\gamma +
  \rm {invisible}$}},
  \href{https://doi.org/10.1103/PhysRevD.101.112005}{\emph{Phys. Rev. D}
  {\bfseries 101} (2020) 112005}
  [\href{https://arxiv.org/abs/2003.05594}{{\ttfamily 2003.05594}}].

\bibitem{BaBar:2010eww}
{\scshape BaBar} collaboration, \emph{{Search for Production of Invisible Final
  States in Single-Photon Decays of $\Upsilon(1S)$}},
  \href{https://doi.org/10.1103/PhysRevLett.107.021804}{\emph{Phys. Rev. Lett.}
  {\bfseries 107} (2011) 021804}
  [\href{https://arxiv.org/abs/1007.4646}{{\ttfamily 1007.4646}}].

\bibitem{Belle:2018pzt}
{\scshape Belle} collaboration, \emph{{Search for a light $CP$-odd Higgs boson
  and low-mass dark matter at the Belle experiment}},
  \href{https://doi.org/10.1103/PhysRevLett.122.011801}{\emph{Phys. Rev. Lett.}
  {\bfseries 122} (2019) 011801}
  [\href{https://arxiv.org/abs/1809.05222}{{\ttfamily 1809.05222}}].

\bibitem{BaBar:2008aby}
{\scshape BaBar} collaboration, \emph{{Search for Invisible Decays of a Light
  Scalar in Radiative Transitions $\Upsilon_{3S} \to \gamma A^{0}$}},  in
  \emph{{ 34th International Conference on High Energy Physics}}, 7, 2008
  [\href{https://arxiv.org/abs/0808.0017}{{\ttfamily 0808.0017}}].

\bibitem{Knapen:2016moh}
S.~Knapen, T.~Lin, H.K.~Lou and T.~Melia, \emph{{Searching for Axionlike
  Particles with Ultraperipheral Heavy-Ion Collisions}},
  \href{https://doi.org/10.1103/PhysRevLett.118.171801}{\emph{Phys. Rev. Lett.}
  {\bfseries 118} (2017) 171801}
  [\href{https://arxiv.org/abs/1607.06083}{{\ttfamily 1607.06083}}].

\bibitem{OPAL:2002vhf}
{\scshape OPAL} collaboration, \emph{{Multiphoton production in $e^{+}$ $e^{-}$
  collisions at $\sqrt{s}$ = 181 GeV to 209 GeV}},
  \href{https://doi.org/10.1140/epjc/s2002-01074-5}{\emph{Eur. Phys. J. C}
  {\bfseries 26} (2003) 331}
  [\href{https://arxiv.org/abs/hep-ex/0210016}{{\ttfamily hep-ex/0210016}}].

\bibitem{L3:1994shn}
{\scshape L3} collaboration, \emph{{Search for anomalous
  $Z\to\gamma\gamma\gamma$ events at LEP}},
  \href{https://doi.org/10.1016/0370-2693(95)01612-T}{\emph{Phys. Lett. B}
  {\bfseries 345} (1995) 609}.

\bibitem{Belle-II:2020jti}
{\scshape Belle-II} collaboration, \emph{{Search for Axion-Like Particles
  produced in $e^+e^-$ collisions at Belle II}},
  \href{https://doi.org/10.1103/PhysRevLett.125.161806}{\emph{Phys. Rev. Lett.}
  {\bfseries 125} (2020) 161806}
  [\href{https://arxiv.org/abs/2007.13071}{{\ttfamily 2007.13071}}].

\bibitem{CMS:2018erd}
{\scshape CMS} collaboration, \emph{{Evidence for light-by-light scattering and
  searches for axion-like particles in ultraperipheral PbPb collisions at
  $\sqrt{s_\mathrm{NN}} =$ 5.02 TeV}},
  \href{https://doi.org/10.1016/j.physletb.2019.134826}{\emph{Phys. Lett. B}
  {\bfseries 797} (2019) 134826}
  [\href{https://arxiv.org/abs/1810.04602}{{\ttfamily 1810.04602}}].

\bibitem{ATLAS:2020hii}
{\scshape ATLAS} collaboration, \emph{{Measurement of light-by-light scattering
  and search for axion-like particles with 2.2 nb$^{-1}$ of PbPb data with the
  ATLAS detector}}, \href{https://doi.org/10.1007/JHEP03(2021)243}{\emph{JHEP}
  {\bfseries 03} (2021) 243}
  [\href{https://arxiv.org/abs/2008.05355}{{\ttfamily 2008.05355}}].

\bibitem{BESIII:2022rzz}
{\scshape BESIII} collaboration, \emph{{Search for an axion-like particle in
  radiative J/\ensuremath{\psi} decays}},
  \href{https://doi.org/10.1016/j.physletb.2023.137698}{\emph{Phys. Lett. B}
  {\bfseries 838} (2023) 137698}
  [\href{https://arxiv.org/abs/2211.12699}{{\ttfamily 2211.12699}}].

\bibitem{BESIII:2024hdv}
{\scshape BESIII} collaboration, \emph{{Search for diphoton decays of an
  axionlike particle in radiative J/\ensuremath{\psi} decays}},
  \href{https://doi.org/10.1103/PhysRevD.110.L031101}{\emph{Phys. Rev. D}
  {\bfseries 110} (2024) L031101}
  [\href{https://arxiv.org/abs/2404.04640}{{\ttfamily 2404.04640}}].

\bibitem{BESIII:2024rex}
{\scshape BESIII} collaboration, \emph{{Observation of the charmonium decay
  $\eta_c\to\gamma\gamma$ in $J/\psi\to\gamma\eta_c$}},
  \href{https://doi.org/10.1103/PhysRevLett.134.181901}{\emph{Phys. Rev. Lett.}
  {\bfseries 134} (2025) 181901}
  [\href{https://arxiv.org/abs/2412.12998}{{\ttfamily 2412.12998}}].

\bibitem{Merlo:2019anv}
L.~Merlo, F.~Pobbe, S.~Rigolin and O.~Sumensari, \emph{{Revisiting the
  production of ALPs at B-factories}},
  \href{https://doi.org/10.1007/JHEP06(2019)091}{\emph{JHEP} {\bfseries 06}
  (2019) 091} [\href{https://arxiv.org/abs/1905.03259}{{\ttfamily
  1905.03259}}].

\bibitem{DiLuzio:2024jip}
L.~Di~Luzio, A.W.M.~Guerrera, X.~Ponce~D\'\i{}az and S.~Rigolin,
  \emph{{Axion-like particles in radiative quarkonia decays}},
  \href{https://doi.org/10.1007/JHEP06(2024)217}{\emph{JHEP} {\bfseries 06}
  (2024) 217} [\href{https://arxiv.org/abs/2402.12454}{{\ttfamily
  2402.12454}}].

\bibitem{pdg:2024}
{\scshape Particle Data Group} collaboration, \emph{{Review of particle
  physics}}, \href{https://doi.org/10.1103/PhysRevD.110.030001}{\emph{Phys.
  Rev. D} {\bfseries 110} (2024) 030001}.

\bibitem{Becirevic:2013bsa}
D.~Be\v{c}irevi\'c, G.~Duplan\v{c}i\'c, B.~Klajn, B.~Meli\'c and F.~Sanfilippo,
  \emph{{Lattice QCD and QCD sum rule determination of the decay constants of
  $\eta_c$, J/$\psi$ and $h_c$ states}},
  \href{https://doi.org/10.1016/j.nuclphysb.2014.03.024}{\emph{Nucl. Phys. B}
  {\bfseries 883} (2014) 306}
  [\href{https://arxiv.org/abs/1312.2858}{{\ttfamily 1312.2858}}].

\bibitem{Nason:1986tr}
P.~Nason, \emph{{{QCD} Radiative Corrections to $\Upsilon$ Decay Into Scalar
  Plus $\gamma$ and Pseudoscalar Plus $\gamma$}},
  \href{https://doi.org/10.1016/0370-2693(86)90721-5}{\emph{Phys. Lett. B}
  {\bfseries 175} (1986) 223}.

\bibitem{Bauer:2021mvw}
M.~Bauer, M.~Neubert, S.~Renner, M.~Schnubel and A.~Thamm, \emph{{Flavor probes
  of axion-like particles}},
  \href{https://doi.org/10.1007/JHEP09(2022)056}{\emph{JHEP} {\bfseries 09}
  (2022) 056} [\href{https://arxiv.org/abs/2110.10698}{{\ttfamily
  2110.10698}}].

\bibitem{Carmona:2021seb}
A.~Carmona, C.~Scherb and P.~Schwaller, \emph{{Charming ALPs}},
  \href{https://doi.org/10.1007/JHEP08(2021)121}{\emph{JHEP} {\bfseries 08}
  (2021) 121} [\href{https://arxiv.org/abs/2101.07803}{{\ttfamily
  2101.07803}}].

\bibitem{Colquhoun:2025xlx}
{\scshape HPQCD} collaboration, \emph{{Constraints on axion-like particles
  using lattice QCD calculations of the rate for $J/\psi \to \gamma a$}},
  \href{https://arxiv.org/abs/2502.06721}{{\ttfamily 2502.06721}}.

\bibitem{Dolan:2017osp}
M.J.~Dolan, T.~Ferber, C.~Hearty, F.~Kahlhoefer and K.~Schmidt-Hoberg,
  \emph{{Revised constraints and Belle II sensitivity for visible and invisible
  axion-like particles}},
  \href{https://doi.org/10.1007/JHEP12(2017)094}{\emph{JHEP} {\bfseries 12}
  (2017) 094} [\href{https://arxiv.org/abs/1709.00009}{{\ttfamily
  1709.00009}}].

\bibitem{Allen:2024ndv}
S.~Allen, A.~Blackburn, O.~Cardenas, Z.~Messenger, N.H.~Nguyen and B.~Shuve,
  \emph{{Electroweak axion portal to dark matter}},
  \href{https://doi.org/10.1103/PhysRevD.110.095010}{\emph{Phys. Rev. D}
  {\bfseries 110} (2024) 095010}
  [\href{https://arxiv.org/abs/2405.02403}{{\ttfamily 2405.02403}}].

\bibitem{Kaneta:2016wvf}
K.~Kaneta, H.S.~Lee and S.~Yun, \emph{{Portal Connecting Dark Photons and
  Axions}}, \href{https://doi.org/10.1103/PhysRevLett.118.101802}{\emph{Phys.
  Rev. Lett.} {\bfseries 118} (2017) 101802}
  [\href{https://arxiv.org/abs/1611.01466}{{\ttfamily 1611.01466}}].

\bibitem{BaBar:2017tiz}
{\scshape BaBar} collaboration, \emph{{Search for Invisible Decays of a Dark
  Photon Produced in ${e}^{+}{e}^{-}$ Collisions at BaBar}},
  \href{https://doi.org/10.1103/PhysRevLett.119.131804}{\emph{Phys. Rev. Lett.}
  {\bfseries 119} (2017) 131804}
  [\href{https://arxiv.org/abs/1702.03327}{{\ttfamily 1702.03327}}].

\bibitem{Achasov:2023gey}
M.~Achasov et~al., \emph{{STCF conceptual design report (Volume 1): Physics
  $\&$ detector}},
  \href{https://doi.org/10.1007/s11467-023-1333-z}{\emph{Front. Phys.
  (Beijing)} {\bfseries 19} (2024) 14701}
  [\href{https://arxiv.org/abs/2303.15790}{{\ttfamily 2303.15790}}].

\bibitem{Belle-II:2010dht}
{\scshape Belle-II} collaboration, \emph{{Belle II Technical Design Report}},
  \href{https://arxiv.org/abs/1011.0352}{{\ttfamily 1011.0352}}.

\bibitem{BESIII:2021cxx}
{\scshape BESIII} collaboration, \emph{{Number of $J/\psi$ events at BESIII}},
  \href{https://doi.org/10.1088/1674-1137/ac5c2e}{\emph{Chin. Phys. C}
  {\bfseries 46} (2022) 074001}
  [\href{https://arxiv.org/abs/2111.07571}{{\ttfamily 2111.07571}}].

\bibitem{BESIII:2017tvm}
{\scshape BESIII Collaboration} collaboration, \emph{{Determination of the
  number of $\psi(3686)$ events at BESIII}},
  \href{https://doi.org/10.1088/1674-1137/42/2/023001}{\emph{Chin. Phys. C}
  {\bfseries 42} (2018) 023001}
  [\href{https://arxiv.org/abs/1709.03653}{{\ttfamily 1709.03653}}].

\bibitem{BESIII:2024lks}
{\scshape BESIII Collaboration} collaboration, \emph{{Determination of the
  number of $\psi(3686)$ events taken at BESIII}},
  \href{https://doi.org/10.1088/1674-1137/ad595b}{\emph{Chin. Phys. C}
  {\bfseries 48} (2024) 093001}
  [\href{https://arxiv.org/abs/2403.06766}{{\ttfamily 2403.06766}}].

\bibitem{Holdom:1985ag}
B.~Holdom, \emph{{Two U(1)'s and Epsilon Charge Shifts}},
  \href{https://doi.org/10.1016/0370-2693(86)91377-8}{\emph{Phys. Lett. B}
  {\bfseries 166} (1986) 196}.

\bibitem{Buschmann:2015awa}
M.~Buschmann, J.~Kopp, J.~Liu and P.A.N.~Machado, \emph{{Lepton Jets from
  Radiating Dark Matter}},
  \href{https://doi.org/10.1007/JHEP07(2015)045}{\emph{JHEP} {\bfseries 07}
  (2015) 045} [\href{https://arxiv.org/abs/1505.07459}{{\ttfamily
  1505.07459}}].

\end{thebibliography}\endgroup

\newpage
\end{document}